\documentclass[a4paper,11pt]{article}

\usepackage{amsmath,amssymb,mathtools}
\usepackage{color}
\usepackage{graphicx}
\usepackage{subfigure}
\usepackage{cite}
\usepackage[colorlinks=true,linkcolor=red,citecolor=blue,urlcolor=blue,bookmarks]{hyperref}
\usepackage{multirow,makecell}
\usepackage{textcomp}
\usepackage{wasysym}
\usepackage{ulem}

\usepackage{verbatim}

\usepackage[utf8]{inputenc}
\usepackage[T1]{fontenc}

\usepackage{array}

\usepackage{diagbox}  

\usepackage[text={17.1cm,24.6cm},centering]{geometry} 

\numberwithin{equation}{section}

\def \be {\begin{equation}}
\def \ee {\end{equation}}
\def \ba {\begin{array}}
\def \ea {\end{array}}
\def \bea {\begin{eqnarray}}
\def \eea {\end{eqnarray}}
\def \nn {\nonumber}

\def \a {\alpha}
\def \b {\beta}

\def \G {\Gamma}
\def \d {\delta}
\def \D {\Delta}
\def \e {\epsilon}
\def \ve {\varepsilon}

\def \m {\mu}

\def \s {\sigma}

\def \r {\rho}

\def \vphi {\varphi}

\def \cA {\mathcal A}

\def \cF {\mathcal F}

\def \cT {\mathcal T}

\def \cX {\mathcal X}

\def \p {\partial}

\def \f {\frac}

\def \lt {\left}
\def \rt {\right}

\def \sr {\sqrt}
\def \td {\tilde}

\def \inf {\infty}

\def \lag {\langle}
\def \rag {\rangle}

\def \dd {\mathrm{d}}
\def \ep {\mathrm{e}}
\def \ii {\mathrm{i}}

\def \tr {\textrm{tr}}

\def \and {{~\textrm{and}~}}

\begin{document}

\title{
\textbf{Subsystem fidelity in two-dimensional conformal field theories}
}

\author{
Bin Sui,
Yihao Wang
and
Jiaju Zhang\footnote{Corresponding author: jiajuzhang@tju.edu.cn}
}
\date{}
\maketitle
\vspace{-10mm}
\begin{center}
{\it
Center for Joint Quantum Studies and Department of Physics, School of Science, Tianjin University,\
135 Yaguan Road, Tianjin 300350, China
}
\vspace{10mm}
\end{center}

\begin{abstract}

  We investigate the short-interval expansion of the subsystem fidelity in two-dimensional conformal field theories (2D CFTs) using the operator product expansion (OPE) of twist operators. We obtain universal contributions from general quasiprimary operators valid for arbitrary 2D CFTs, along with specific results in free massless boson and fermion theories. The analytical predictions demonstrate excellent agreement with established analytical results in field theories and numerical calculations in integrable models. Furthermore, we extend the method to holographic CFTs, where subsystem fidelity serves to analyze the distinguishability of black hole microstates through the AdS/CFT correspondence. This work establishes a unified framework for quantifying quantum state distinguishability across various 2D CFTs, bridging quantum information techniques with applications in quantum gravity.

\end{abstract}

\baselineskip 18pt
\thispagestyle{empty}
\newpage

\tableofcontents

\section{Introduction}\label{sectionIntroduction}

The quantitative distinguishability between quantum states plays a fundamental role across multiple domains of theoretical physics. Quantitative measures of distinguishability not only serve as operational tools but also bridge fundamental concepts between quantum information, many-body physics, and quantum gravity. In quantum information theory, it provides a crucial metric for evaluating the precision and reliability of quantum states in various operational tasks, such as quantum metrology and quantum computing, where optimal discrimination between states directly impacts protocol performance \cite{Nielsen:2010oan,Watrous:2018rgz}. Moreover, in the study of quantum thermalization, state distinguishability offers a powerful diagnostic tool to characterize and compare thermalization processes in chaotic versus integrable quantum systems, thereby offering insights into the emergence of statistical mechanics from unitary quantum dynamics \cite{Banuls:2010fkc,Fagotti:2013jzu,Cardy:2014rqa,Caputa:2016yzn}. Furthermore, this concept has been rigorously formulated within quantum field theory \cite{Lashkari:2014yva,Lashkari:2015dia,Zhang:2019wqo,Zhang:2019itb}, where it helps quantify the distinguishability of eigenstates and states under unitary evolution, with significant implications for understanding the structure of the Hilbert space. In the context of the AdS/CFT correspondence \cite{Maldacena:1997re,Witten:1998qj,Gubser:1998bc}, state distinguishability is intimately related to the problem of bulk reconstruction, determining how boundary quantum states encode information about the geometry and fields in the anti-de Sitter bulk spacetime \cite{Jafferis:2015del,Dong:2016eik,Suzuki:2019xdq,Kusuki:2019hcg}.

In this paper, we focus on calculating subsystem fidelity in two-dimensional conformal field theories (2D CFTs). Fidelity measures the similarity between two quantum states. For two density matrices $\rho$ and $\sigma$, it is defined as \cite{Nielsen:2010oan,Watrous:2018rgz}
\be \label{Fdef1}
F(\rho,\sigma) = \tr\sqrt{ \sqrt{\rho} \sigma \sqrt{\rho} }.
\ee
Direct evaluation of this expression with a large Hilbert space is often challenging due to the presence of square roots. Inspired by the replica trick used in entanglement entropy calculations \cite{Holzhey:1994we,Calabrese:2004eu}, several replica-based approaches have been developed to compute fidelity.
In \cite{Lashkari:2014yva}, the following replica expression was used
\be
F(\rho,\sigma) = \lim_{p \to 1/2} \tr \big[ \big( \rho^{\frac{1-p}{2p}} \sigma \rho^{\frac{1-p}{2p}} \big)^p \big].
\ee
A simplified two-parameter replica trick was later used in \cite{Suzuki:2019xdq,Kusuki:2019hcg}
\be
F(\rho,\sigma) = \lim_{m \to 1/2,, n \to 1/2} \tr[ ( \rho^m \sigma \rho^m )^n ].
\ee
A more computationally efficient formula of fidelity was suggested in \cite{Baldwin:2022cjb}
\be \label{Fdef2}
F(\rho,\sigma) = \tr\sqrt{\rho \sigma}.
\ee
More recently, a refined replica trick was introduced \cite{Parez:2022sgc,Xiao:2023hnh}
\be \label{reptri2}
F(\rho,\sigma) = \lim_{p \to 1/2} \tr[ (\rho \sigma)^p ].
\ee
In this work, we adopt this last replica trick to evaluate subsystem fidelity in 2D CFTs.

Although not immediately apparent, the new definition of fidelity (\ref{Fdef2}), proposed in \cite{Baldwin:2022cjb}, is in fact equivalent to the conventional expression (\ref{Fdef1}). This equivalence can be understood via the replica trick. To evaluate the standard fidelity (\ref{Fdef1}), one employs the identity
\be \label{reptri1}
F(\rho,\sigma) = \lim_{p\to1/2} \tr \big[ (\sqrt{\rho} \sigma \sqrt{\rho})^p \big].
\ee
Meanwhile, for any nonnegative integer $p$, the identity \(\tr[(\sqrt{\rho} \sigma \sqrt{\rho})^p] = \tr[(\rho \sigma)^p]\) always holds. As a result, the replica trick in (\ref{reptri2}) yields the same limit as (\ref{reptri1}), confirming that both definitions coincide. Alternatively, one may observe that for any matrices $X$ and $Y$ of compatible dimensions and any nonnegative integer $p$, the equality \(\tr[(XY)^p] = \tr[(YX)^p]\) implies that $XY$ and $YX$ share the same nonvanishing eigenvalues. In particular, $\sqrt{X} Y \sqrt{X}$ and $XY$ are isospectral, provided the relevant matrix square roots exist. Consequently, their square roots, $\sqrt{\sqrt{X} Y \sqrt{X}}$ and $\sqrt{XY}$, also have identical eigenvalues. This reasoning again shows that definitions (\ref{Fdef1}) and (\ref{Fdef2}) are equivalent, regardless of whether $\rho$ and $\sigma$ commute.

We study subsystem fidelity in general 2D CFTs using twist operators \cite{Calabrese:2004eu,Cardy:2007mb,Calabrese:2009qy} and their operator product expansion (OPE) \cite{Headrick:2010zt,Calabrese:2010he,Rajabpour:2011pt,Chen:2013kpa,Chen:2013dxa,Lin:2016dxa,Chen:2016lbu,He:2017vyf,Basu:2017kzo,%
He:2017txy,Ruggiero:2018hyl,Guo:2018fye,Zhang:2019wqo,Zhang:2019itb}. By evaluating the short-interval expansion of subsystem fidelity via the OPE of twist operators, we show that it receives contributions from various quasiprimary operators ordered by their scaling dimensions. We derive universal contributions from general quasiprimary operators in arbitrary 2D CFTs, as well as specific contributions from particular operators in free boson and free fermion theories. Our results in these solvable models agree with existing analytical and numerical calculations. Furthermore, we extend the method to 2D holographic CFTs, where through the AdS$_3$/CFT$_2$ correspondence \cite{Brown:1986nw}, we analyze the perturbative distinguishability of black hole microstates.

The paper is organized as follows. In Section~\ref{sectionExpansion}, we outline the general method for computing subsystem fidelity using the short-interval expansion of twist operator OPEs, including both universal and model-specific contributions. In Sections~\ref{sectionBoson} and~\ref{sectionFermion}, we validate the approach by calculating subsystem fidelity between low-lying eigenstates in the 2D massless free boson and fermion theories, respectively. Section~\ref{sectionHolographic} applies the framework to holographic CFTs, exploring the distinguishability of black hole microstates. We conclude in Section~\ref{sectionConclusion} with a summary and discussion. Additional technical details on the analytical continuations used in the fidelity calculations are provided in Appendix~\ref{appendixAnalytical}.

\section{General method} \label{sectionExpansion}

In this section, we outline the general method for calculating the short-interval expansion of subsystem fidelity using the OPE of twist operators in 2D CFTs. We begin by reviewing relevant basic concepts, then apply the replica trick to derive subsystem fidelity expressed in terms of contributions from various quasiprimary operators in the replicated CFT. We discuss several universal contributions to subsystem fidelity in general 2D CFTs and two specific types of particular contributions in free massless boson and fermion theories.

\subsection{Relevant basics}

In this subsection, we review essential background concepts of 2D CFTs. For more details, see \cite{Ginsparg:1988ui,DiFrancesco:1997nk,Blumenhagen:2009zz}.

From global conformal symmetry, all independent operators in a general 2D CFT can be classified as primary operators and their descendants, while from the global SL$(2,C)$ conformal symmetry, all operators can be classified as quasiprimary operators and their derivatives. Under a general local conformal transformation $z \to f(z)$, a primary operator $\phi$ with conformal weights $(h_\phi,\bar h_\phi)$ transforms as
\be
\phi(z,\bar z) = f'(z)^{h_\phi} \bar f'(\bar z)^{\bar h_\phi} \phi(f(z),\bar f(\bar z)),
\ee
while a quasiprimary operator $\phi$ transforms as
\be \label{phizbarz}
\phi(z,\bar z) = f'(z)^{h_\phi} \bar f'(\bar z)^{\bar h_\phi} \phi(f(z),\bar f(\bar z)) + \cdots,
\ee
where $\cdots$ denotes terms containing the Schwarz derivative
\be
s(z) = \f{f'''(z)}{f'(z)} - \f32 \Big( \f{f''(z)}{f'(z)} \Big)^2,
\ee
and/or its derivatives. From the conformal weights $(h_\phi,\bar h_\phi)$, one obtains the scaling dimension $\D_\phi=h_\phi+\bar h_\phi$ and spin $s_\phi=h_\phi-\bar h_\phi$. The operator $\phi$ is bosonic when $s_\phi$ is an integer and fermionic when $s_\phi$ is half-integer. In both cases, $\ii^{8s_\phi}=1$.

The correlation function for two quasiprimary operators on a complex plane $C$ is
\be
\lag \phi(z_1,\bar z_1) \psi(z_2,\bar z_2) \rag_C = \f{\a_\phi \d_{\phi\psi}}{z_{12}^{2h_\phi}\bar z_{12}^{2\bar h_\phi}},
\ee
where $\a_\phi$ relates to the normalization of operator $\phi$, all quasiprimary operators have been orthogonalized, and we use $z_{12} \equiv z_1-z_2$ and $\bar z_{12} \equiv \bar z_1-\bar z_2$.
The correlation function of three quasiprimary operators on a plane is
\be
\lag \phi(z_1,\bar z_1) \psi(z_2,\bar z_2) \chi(z_3,\bar z_3) \rag_C
= \f{C_{\phi\psi\chi}}
{
z_{12}^{h_\phi+h_\psi-h_\chi} z_{13}^{h_\phi+h_\chi-h_\psi} z_{23}^{h_\psi+h_\chi-h_\phi}
\bar z_{12}^{\bar h_\phi+\bar h_\psi-\bar h_\chi}
\bar z_{13}^{\bar h_\phi+\bar h_\chi-\bar h_\psi}
\bar z_{23}^{\bar h_\psi+\bar h_\chi-\bar h_\phi}
},
\ee
where $C_{\phi\psi\chi}$ is the structure constant.

Typical quasiprimary operators include the stress tensor $T$ with conformal weights $(2,0)$ and normalization $\a_T=\f{c}{2}$, where $c$ is the central charge of the 2D CFT, and the operator
\be
\cA = (TT) - \f{3}{10} \p^2 T,
\ee
with conformal weights $(4,0)$ and normalization
\be
\a_{\cA}=\f{c(5c+22)}{10}.
\ee
Similar considerations apply to the anti-holomorphic stress tensor $\bar T$ with conformal weights $(0,2)$ and the operator $\bar\cA = (\bar T\bar T) - \f{3}{10} \bar\p^2 \bar T$ with conformal weights $(0,4)$.

A pure state $|\phi\rag$ in a 2D CFT on a cylinder with coordinate $w$ and spatial period $w \sim w+L$ corresponds to inserting an operator $\phi(z,\bar z)$ at the origin on a plane with coordinate $z=\ep^{\f{2\pi\ii w}{L}}$
\be
|\phi\rag = \phi(0,0) |G\rag,
\ee
with $|G\rag$ denoting the ground state.
When $\phi$ is a primary operator, we have the expectation values
\be \label{TphiTbarphi}
\lag T \rag_\phi = \frac{\pi^2 (c-24 h_\phi)}{6 L^2}, ~~
\lag \cA \rag_\phi = \frac{\pi^4 [ c (5 c+22) -240 (c+2) h_\phi +2880 h_\phi^2 ]}{180 L^4}.
\ee
It is similar for $\lag \bar T \rag_\phi$ and $\lag \bar \cA \rag_\phi$.

We also consider the thermal state
\be
\r_\b = \f{\ep^{-\b H}}{Z(\b)},
\ee
where $\b$ is the inverse temperature, $H$ is the Hamiltonian, and $Z(\b)=\tr (\ep^{-\b H})$ is the partition function.
Useful thermal state expectation values are
\be
\lag T \rag_\b = -\f{\pi^2c}{6\b^2}, ~~
\lag \cA \rag_\b = \f{\pi^4c(5c+22)}{180\b^4}.
\ee
It is the same for $\lag \bar T \rag_\b$ and $\lag \bar \cA \rag_\b$.

\subsection{OPE of twist operators}

We consider one interval $A=[0,\ell]$ on a circle of length $L$ in two general states $\r$ and $\s$ in a general 2D CFT. The subsystem fidelity can be calculated as \cite{Baldwin:2022cjb}
\be
F(\r_A,\s_A) = \tr_A \sqrt{\r_A \s_A}.
\ee
We define the ``generalized fidelity'' as%
\footnote{%
The ``R\'enyi fidelity'' was introduced in \cite{Parez:2022sgc} as
\[ F_p(\r,\s)=\f{\tr[(\r\s)^p]}{\sqrt{\tr(\r^{2p})\tr(\r^{2p})}}. \]
The special $p=1$ case of the R\'enyi fidelity was proposed in \cite{Wang:2008gcy} and has since been applied to problems in thermalization and revival in two-dimensional conformal field theories \cite{Cardy:2014rqa},  bulk reconstruction in AdS/CFT correspondence \cite{Kusuki:2019hcg}, and non-equilibrium evolution after a quantum quench \cite{Parez:2025wtd,Parez:2025psa}.%
}%
\be
F_p(\r_A,\s_A) = \tr_A [ (\r_A \s_A)^p ],
\ee
and obtain fidelity from the limit \cite{Parez:2022sgc,Xiao:2023hnh}
\be
F(\r_A,\s_A) = \lim_{p \to 1/2} F_p(\r_A , \s_A).
\ee
The replica approach proceeds as follows: we first evaluate the generalized fidelity $F_p(\r_A,\s_A)$ for general positive integer values of $p=1,2,\cdots$, corresponding to positive even integer values of $n=2p=2,4,\cdots$, and then apply analytical continuation $p\to1/2$ (i.e., $n\to1$) to obtain the fidelity.

In this paper, we only consider translationally invariant states. Using twist operators $\cT$ and $\td \cT$ \cite{Calabrese:2004eu,Cardy:2007mb,Calabrese:2009qy} and their OPE \cite{Headrick:2010zt,Calabrese:2010he,Rajabpour:2011pt,Chen:2013kpa,Lin:2016dxa,Chen:2016lbu,He:2017vyf,Basu:2017kzo,%
He:2017txy,Guo:2018fye,Ruggiero:2018hyl,Zhang:2019wqo,Zhang:2019itb}, we obtain the short interval expansion
\be \label{TellelltdTellell}
\lag \cT(\ell,\ell) \td \cT(0,0) \rag_{\r\otimes\s\otimes\r\otimes\s\cdots}
= c_n \Big(\f{\ell}{\e}\Big)^{-2(h_n+\bar h_n)}
\Big( 1+\sum_K \ell^{\D_K} d_K \lag \Phi_K \rag_{\r\otimes\s\otimes\r\otimes\s\cdots} \Big).
\ee
The twist operators $\cT$ and $\td\cT$ are primary operators in the replicated $n$-fold CFT, which we denote as CFT$^n$, and their conformal weights are \cite{Calabrese:2004eu,Calabrese:2009qy}
\be
h_n=\bar h_n=\f{c(n^2-1)}{24n},
\ee
where $c$ is the central charge of the original single copy CFT.
The factor $c_n$ relates to the normalization of twist operators, satisfying $\lim_{n\to1}c_n=1$, and $\e$ is the UV cutoff.
Due to translational invariance, we only need to consider CFT$^n$ non-identity quasiprimary operators $\Phi_K$ that are direct products of quasiprimary operators in different replicas
\be \label{PhiK}
\Phi_K = \cX_1^{j_1} \cX_2^{j_2} \cdots \cX_k^{j_k}.
\ee
The set $\{\cX_1,\cdots,\cX_k\}$ consists of non-identity quasiprimary operators in the original single copy of the CFT.
Here $0 \leq j_1,j_2,\cdots,j_k\leq n-1$ are replica indices taking appropriate values to avoid undercounting or overcounting of CFT$^n$ quasiprimary operators. Also because of translational invariance, terms involving the derivatives of  the CFT\(^n\) quasiprimary operators \(\Phi_K\) do not appear in (\ref{TellelltdTellell}).

We obtain the expansion of generalized fidelity
\be
F_p(\r_A,\s_A) = c_n \Big(\f{\ell}{\e}\Big)^{-2(h_n+\bar h_n)}
\bigg(
1 + \sum_{k=1}^n
\sum_{\{\cX_1,\cdots,\cX_k\}}
\ell^{\D_{\cX_1}+\cdots+\D_{\cX_k}}
\cF_{\cX_1 \cdots \cX_k}^{(p)}
\bigg),
\ee
where we define
\be
\cF_{\cX_1 \cdots \cX_k}^{(p)} \equiv \sum_{j_1,\cdots,j_k}
d_{\cX_1\cdots\cX_k}^{j_1\cdots j_k}
\lag \cX_1 \rag_{j_1}\cdots \lag \cX_k \rag_{j_k},
\ee
and the expectation value
\be
\lag \cX \rag_j \equiv \lt\{
\ba{cl}
\lag \cX \rag_\r & j \rm{~is~even} \\
\lag \cX \rag_\s & j \rm{~is~odd}
\ea
\rt.\!\!\!.
\ee

Note that $\cF_{\cX_1 \cdots \cX_k}^{(p)}$ depends on states $\r,\s$, which we omit for conciseness.
For $k=1$, there is always $d_\cX=O(n-1)$ \cite{Guo:2018fye}, so $\cF_{\cX}^{(1/2)}=0$.
We obtain the short interval expansion of subsystem fidelity
\be \label{FrAsA}
F(\r_A,\s_A) = 1 + \sum_{k=2}^{+\inf}
\sum_{\cX_1,\cdots,\cX_k}
\ell^{\D_{\cX_1}+\cdots+\D_{\cX_k}}
\cF_{\cX_1 \cdots \cX_k}^{(1/2)}.
\ee
This expansion provides a systematic way to compute subsystem fidelity order by order in the interval length $\ell$, with contributions from various combinations of quasiprimary operators.%

Formula (\ref{TellelltdTellell}) applies only to translationally invariant states; generalization to inhomogeneous states is possible but more complicated. As with entanglement entropy, fidelity in such cases depends on position of the subsystem.
From the OPE of twist operators we have \cite{Chen:2013kpa,Chen:2013dxa}
\be
\lag \cT(\ell,\ell) \td \cT(0,0) \rag_{\r\otimes\s\otimes\r\otimes\s\cdots}
= c_n \Big(\f{\ell}{\e}\Big)^{-2(h_n+\bar h_n)}
\Big( 1+ \sum_K \sum_{r,s=0}^{+\inf} \f{a_K^r}{r!} \f{\bar a_K^s}{s!} \ell^{\D_K+r+s} d_K \lag \p^r \bar \p^s \Phi_K(0,0) \rag_{\r\otimes\s\otimes\r\otimes\s\cdots} \Big),
\ee
where the coefficients are expressed through binomial coefficients as
\be
a_K^r\equiv \f{C_{h_K+r-1}^r}{C_{2h_K+r-1}^r}, ~~
\bar a_K^s\equiv\f{C_{\bar h_K+s-1}^s}{C_{2\bar h_K+s-1}^s}.
\ee
Moreover, the CFT$^n$ quasiprimary operators $\Phi_K$ are not restricted to the form (\ref{PhiK}); more general terms with derivatives must be included, such as \cite{Zhang:2016rja}
\bea
&& \cX_{j_1}\ii\p \cX_{j_2}-\ii\p \cX_{j_1}\cX_{j_2}, \nn\\
&& \cX_{j_1}\p\bar\p \cX_{j_2}+\p\bar\p \cX_{j_1} \cX_{j_2}-\p \cX_{j_1}\bar\p \cX_{j_2}-\bar\p \cX_{j_1}\p \cX_{j_2}, \nn\\
&& \p \cX_{j_1}\p \cX_{j_2}-\f{h_\cX}{2h_\cX+1} \lt( \cX_{j_1}\p^2 \cX_{j_2} + \p^2 \cX_{j_1} \cX_{j_2} \rt).
\eea
Interestingly, even for inhomogeneous states, the terms with derivatives do not modify the leading order in the short-interval expansion, and the result from subsection \ref{subsectionXX} remain dominant.

\subsection{Universal contributions from $\cX\cX$} \label{subsectionXX}

In this subsection, we derive the universal contributions to subsystem fidelity from pairs of identical quasiprimary operators $\{\cX,\cX\}$. We have a quasiprimary operator $\cX$ with normalization $\a_\cX$, scaling dimension $\D_\cX$, and spin $s_\cX$. Note that $\cX$ needs not be primary.

The OPE coefficient for the CFT$^n$ operator $\cX_{j_1}\cX_{j_2}$ with $0 \leq j_1 < j_2 \leq n-1$ is given by \cite{Calabrese:2010he,Chen:2014kja,Zhang:2016rja}
\be \label{dXXj1j2}
d_{\cX\cX}^{j_1j_2} = \f{\ii^{2s_\cX}}{(2n)^{2\D_\cX}\a_\cX} \f{1}{\big| \sin\f{\pi j_{12}}{n} \big|^{2\D_\cX} } + O(n-1),
\ee
with shorthand $j_{12}\equiv j_1-j_2$.
The contribution from $\cX_{j_1}\cX_{j_2}$ is the $p\to\f12$ limit (i.e., $n\to1$ limit) of
\be \label{cFpcXcX}
\cF^{(p)}_{\cX\cX} = \sum_{0 \leq j_1 < j_2 \leq n-1} d_{\cX\cX}^{j_1j_2} \lag \cX \rag_{j_1} \lag \cX \rag_{j_2},
\ee
where $j_{12}\equiv j_1-j_2$. For non-primary quasiprimary operators $\cX$, the $\cdots$ terms in the conformal transformation (\ref{phizbarz}) produce $O(n-1)$ contributions to $d_{\cX\cX}^{j_1j_2}$ in (\ref{dXXj1j2}). These contributions remain of order $O(n-1)$ after summing over replica indices $j_1,j_2$, vanish in the $n\to1$ limit, and thus do not affect fidelity.

To evaluate the sum in (\ref{cFpcXcX}), we separate the replica indices into even and odd values. Denoting even integers $a=0,2,\cdots,2p-2$ and odd integers $b=1,3,\cdots,2p-1$, we get
\be
\cF^{(p)}_{\cX\cX} = ( \lag \cX \rag_\r^2 + \lag \cX \rag_\s^2 ) \sum_{a_1<a_2} d_{\cX\cX}^{a_1 a_2}
+ \lag \cX \rag_\r \lag \cX \rag_\s \sum_{a,b} d_{\cX\cX}^{ab}.
\ee
Noting the identity
\be
2 \sum_{a_1<a_2} d_{\cX\cX}^{a_1 a_2} + \sum_{a,b} d_{\cX\cX}^{ab} = O(n-1),
\ee
we can simplify the expression to
\be
\cF^{(p)}_{\cX\cX} = \f{\ii^{2s\cX}}{2^{2\D_\cX}\a_\cX} ( \lag\cX\rag_\r - \lag\cX\rag_\s )^2 G_1(p,\D_\cX) + O(n-1),
\ee
where the function $G_1(p,\D_\cX)$ is defined in (\ref{G1pD}) and encapsulates the sum over even replica indices.
Using the analytical continuation (\ref{G112D}), we obtain the universal contributions from $\cX\cX$ to subsystem fidelity
\be \label{F12XX}
\cF^{(1/2)}_{\cX\cX} = - \f{\G(\D_\cX+\f12)}{2^{2\D_\cX+3}\sqrt{\pi}\G(\D_\cX+1)}
\f{\ii^{2s_\cX}( \lag\cX\rag_\r - \lag\cX\rag_\s )^2}{\a_\cX},
\ee
as reported in \cite{Zhang:2025vaq}.

For $\cF^{(1/2)}_{\cX\cX}$ to be nonvanishing, the operator $\cX$ must be bosonic, meaning the spin $s_\cX$ is an integer, satisfying $\ii^{4s_\cX}=1$. This result shows that the leading contribution from pairs of identical operators depends quadratically on the difference of their expectation values in the two states.

\subsection{Universal contributions from $\cX\cX\cX$}

We now consider contributions from triple products of identical quasiprimary operators $\cX\cX\cX$. The OPE coefficient for the CFT$^n$ quasiprimary operator $\cX_{j_1}\cX_{j_2}\cX_{j_3}$ with $0 \leq j_1 < j_2 <j_3 \leq n-1$ is \cite{Calabrese:2010he,Chen:2014kja,Zhang:2016rja}
\be
d_{\cX\cX\cX}^{j_1j_2j_3} = \f{\ii^{s_\cX} C_{\cX\cX\cX}}{(2n)^{3\D_\cX}\a_\cX^3}
\f{1}{\big| \sin\f{\pi j_{12}}{n} \sin\f{\pi j_{13}}{n} \sin\f{\pi j_{23}}{n} \big|^{\D_\cX}}
+ O(n-1).
\ee
The structure constant $C_{\cX\cX\cX}$ vanishes unless the spin $s_\cX$ is an even integer \cite{Blumenhagen:2009zz}.
The contributions from $\cX_{j_1}\cX_{j_2}\cX_{j_3}$ to fidelity correspond to the $n\to1$ limit of
\be
\cF^{(p)}_{\cX\cX\cX} = \sum_{0 \leq j_1 < j_2 < j_3 \leq n-1} d_{\cX\cX\cX}^{j_1j_2j_3}
\lag \cX \rag_{j_1} \lag \cX \rag_{j_2}\lag \cX \rag_{j_3}.
\ee

Separating the sum into contributions from different combinations of even and odd replica indices, we obtain
\be
\cF^{(p)}_{\cX\cX\cX} = ( \lag \cX \rag_\r^3 + \lag \cX \rag_\s^3 ) \sum_{a_1<a_2<a_3} d_{\cX\cX\cX}^{a_1 a_2 a_3}
+ ( \lag \cX \rag_\r + \lag \cX \rag_\s )\lag \cX \rag_\r \lag \cX \rag_\s
\sum_{a_1<a_2,b} d_{\cX\cX\cX}^{a_1 a_2 b}.
\ee
Using the identity
\be
\sum_{a_1<a_2<a_3} d_{\cX\cX\cX}^{a_1 a_2 a_3} + \sum_{a_1<a_2,b} d_{\cX\cX\cX}^{a_1 a_2 b} = O(n-1),
\ee
we simplify to
\be
\cF^{(p)}_{\cX\cX\cX} = \f{\ii^{s\cX} C_{\cX\cX\cX}}{2^{3\D_\cX}\a_\cX^3}
( \lag \cX \rag_\r - \lag \cX \rag_\s )^2 ( \lag \cX \rag_\r + \lag \cX \rag_\s )
G_2(p,\D_\cX) + O(n-1),
\ee
where the function $G_2(p,\D_\cX)$ is defined in (\ref{G2pD}).
Using the analytical continuation (\ref{G212D}), we obtain the universal contributions from $\cX\cX\cX$ to subsystem fidelity
\be \label{F12XXX}
\cF^{(1/2)}_{\cX\cX\cX} = \f{\G(\D_\cX+\f12)^2}
{2^{2\D_\cX+4}\pi\G(\f{\D_\cX}{2}+1)\G(\f{3\D_\cX}{2}+1)}
\f{\ii^{s_\cX}C_{\cX\cX\cX}( \lag\cX\rag_\r - \lag\cX\rag_\s )^2( \lag\cX\rag_\r + \lag\cX\rag_\s )}
{\a_\cX^3}.
\ee

For $\cF^{(1/2)}_{\cX\cX\cX}$ to be nonvanishing, $s_\cX$ must be an even integer, satisfying $\ii^{2s_\cX}=1$. This cubic contribution provides the sub-leading order correction to the fidelity beyond the quadratic terms.

\subsection{Universal contributions from $T\cX\cX$}

We now examine contributions involving the stress tensor $T$ combined with two identical quasiprimary operators $\cX$. The OPE coefficient for the operator $T_{j_1}\cX_{j_2}\cX_{j_3}$ with $j_1\neq j_2$, $j_1\neq j_3$, $j_2<j_3$ is \cite{Chen:2014kja,Zhang:2016rja}
\be
d_{T\cX\cX}^{j_1j_2j_3} = - \f{2\ii^{2s_\cX}h_\cX}{(2n)^{2\D_\cX+2} c \a_\cX}
\f{1}{\big( \sin\f{\pi j_{12}}{n} \sin\f{\pi j_{13}}{n} \big)^2 \big| \sin\f{\pi j_{23}}{n} \big|^{2\D_\cX-2}} + O(n-1).
\ee
After summing over replica indices and separating contributions, we obtain
\bea
&& \cF_{T\cX\cX}^{(p)} = - \f{\ii^{2s_\cX}h_\cX}{2^{2\D_\cX+1} c \a_\cX}
\big[
( \lag T \rag_\r \lag \cX \rag_\r^2 + \lag T \rag_\s \lag \cX \rag_\s^2 ) G_3(p,\D_\cX)
+ ( \lag T \rag_\r \lag \cX \rag_\s^2 + \lag T \rag_\s \lag \cX \rag_\r^2 ) G_4(p,\D_\cX) \nn\\
&& \phantom{\cF_{T\cX\cX}^{(p)} =}
+ ( \lag T \rag_\r + \lag T \rag_\s ) \lag \cX \rag_\r \lag \cX \rag_\s G_5(p,\D_\cX)
\big]
+O(n-1),
\eea
where $G_3(p,\D_\cX)$, $G_4(p,\D_\cX)$, and $G_5(p,\D_\cX)$ are defined in (\ref{G3pD}), (\ref{G4pD}), and (\ref{G5pD}), respectively.
Using the analytical continuations (\ref{G312D}), (\ref{G412D}), and (\ref{G512D}), we obtain the universal contributions of $T\cX\cX$ to subsystem fidelity
\bea \label{F12TXX}
&& \cF_{T\cX\cX}^{(1/2)} = \f{h_\cX\G(\D_\cX+\f12)}{2^{2\D_\cX+5} \sqrt{\pi} c \G(\D_\cX+2)}
\f{\ii^{2s_\cX}}{\a_\cX}
(\lag \cX \rag_\r-\lag \cX \rag_\s)
\big[
( \lag T \rag_\r + \lag T \rag_\s ) ( \lag \cX \rag_\r - \lag \cX \rag_\s ) \nn\\
&& \phantom{\cF_{T\cX\cX}^{(1/2)} =}
- \D_\cX (
5 \lag T \rag_\r \lag \cX \rag_\r
+ \lag T \rag_\r \lag \cX \rag_\s
- \lag T \rag_\s \lag \cX \rag_\r
-5 \lag T \rag_\s \lag \cX \rag_\s
)
\big].
\eea

Similarly, we obtain contributions of $\bar T\cX\cX$ to subsystem fidelity
\bea \label{F12TbarXX}
&& \cF_{\bar T\cX\cX}^{(1/2)} = \f{\bar h_\cX\G(\D_\cX+\f12)}{2^{2\D_\cX+5} \sqrt{\pi} c \G(\D_\cX+2)}
\f{\ii^{2s_\cX}}{\a_\cX}
(\lag \cX \rag_\r-\lag \cX \rag_\s)
\big[
( \lag \bar T \rag_\r + \lag \bar T \rag_\s ) ( \lag \cX \rag_\r - \lag \cX \rag_\s ) \nn\\
&& \phantom{\cF_{T\cX\cX}^{(1/2)} =}
- \D_\cX (
5 \lag \bar T \rag_\r \lag \cX \rag_\r
+ \lag \bar T \rag_\r \lag \cX \rag_\s
- \lag \bar T \rag_\s \lag \cX \rag_\r
-5 \lag \bar T \rag_\s \lag \cX \rag_\s
)
\big].
\eea

These mixed contributions involving the stress tensor provide important corrections that depend on both the quasiprimary operator $\cX$ and the energy in the two states.

\subsection{Contributions from $JJJJ$ in free massless boson theory}

In free massless boson theory, there is the current operator $J$, which is primary and has normalization $\a_J=1$ and conformal weights $(1,0)$. The OPE coefficient of CFT$^n$ primary operator $J_{j_1}J_{j_2}J_{j_3}J_{j_4}$ with $0\leq j_1<j_2<j_3<j_4\leq n-1$ is \cite{Li:2016qbo}
\be
d_{JJJJ}^{j_1j_2j_3j_4} = \f{1}{(2n)^4} \bigg[
\f{1}{\big( \sin\f{\pi j_{12}}{n} \sin\f{\pi j_{34}}{n} \big)^2}
+\f{1}{\big( \sin\f{\pi j_{13}}{n} \sin\f{\pi j_{24}}{n} \big)^2}
+\f{1}{\big( \sin\f{\pi j_{14}}{n} \sin\f{\pi j_{23}}{n} \big)^2}
\bigg].
\ee
After summing over replica indices and organizing terms by their dependence on the expectation values, we obtain
\be
\cF_{JJJJ}^{(p)} = \f{1}{16} \big[
( \lag J \rag_\r^4 + \lag J \rag_\s^4 ) G_6(p)
+ ( \lag J \rag_\r^2 + \lag J \rag_\s^2 ) \lag J \rag_\r \lag J \rag_\s G_7(p)
+ \lag J \rag_\r^2 \lag J \rag_\s^2 G_8(p)
\big]
+O(n-1),
\ee
where $G_6(p)$, $G_7(p)$, and $G_8(p)$ are defined in (\ref{G6p}), (\ref{G7p}), and (\ref{G8p}), respectively.
Using the analytical continuations (\ref{G67812}), we obtain contributions from $JJJJ$
\be \label{F12JJJJ}
\cF_{JJJJ}^{(1/2)} = - \f{1}{8192} ( \lag J \rag_\r - \lag J \rag_\s )^2
( 19 \lag J \rag_\r^2 + 26 \lag J \rag_\r \lag J \rag_\s + 19\lag J \rag_\s^2 ).
\ee

The operator $\bar J$ has normalization $\a_{\bar J}=1$ and conformal weights $(0,1)$.
Similarly, we obtain contributions from $\bar J\bar J\bar J\bar J$:
\be \label{F12JbarJbarJbarJbar}
\cF_{\bar J\bar J\bar J\bar J}^{(1/2)} = - \f{1}{8192} ( \lag \bar J \rag_\r - \lag \bar J \rag_\s )^2
( 19 \lag \bar J \rag_\r^2
+ 26 \lag \bar J \rag_\r \lag \bar J \rag_\s
+ 19\lag \bar J \rag_\s^2 ).
\ee

These quartic contributions provide higher-order corrections specific to the free boson theory and demonstrate in part how current operators contribute to subsystem fidelity.

\subsection{Contributions from $\varepsilon\varepsilon\varepsilon\varepsilon$ in free massless fermion theory}

In free massless fermion theory, the energy operator $\ve$ is primary and has normalization $\a_\ve=1$ and conformal weights $(h,\bar h)=(\f12,\f12)$. From the correlation function on the plane
\be
\lag \ve(z_1,\bar z_1) \ve(z_2,\bar z_2) \ve(z_3,\bar z_3) \ve(z_4,\bar z_4) \rag_C
= \Big| \f{1}{z_{12}z_{34}}
- \f{1}{z_{13}z_{24}}
+ \f{1}{z_{14}z_{23}} \Big|^2,
\ee
we obtain the OPE coefficient of CFT$^n$ primary operator $\ve_{j_1}\ve_{j_2}\ve_{j_3}\ve_{j_4}$ with $0\leq j_1<j_2<j_3<j_4\leq n-1$
\be
d_{\ve\ve\ve\ve}^{j_1j_2j_3j_4} = \f{1}{(2n)^4} \bigg(
\f{1}{\sin\f{\pi j_{12}}{n} \sin\f{\pi j_{34}}{n}}
-\f{1}{\sin\f{\pi j_{13}}{n} \sin\f{\pi j_{24}}{n}}
+\f{1}{\sin\f{\pi j_{14}}{n} \sin\f{\pi j_{23}}{n}}
\bigg)^2.
\ee
The summation over replica indices yields
\be
\cF_{\ve\ve\ve\ve}^{(p)} = \f{1}{16} \big[
( \lag \ve \rag_\r^4 + \lag \ve \rag_\s^4 ) G_9(p)
+ ( \lag \ve \rag_\r^2 + \lag \ve \rag_\s^2 ) \lag \ve \rag_\r \lag \ve \rag_\s G_{10}(p)
+ \lag \ve \rag_\r^2 \lag \ve \rag_\s^2 G_{11}(p)
\big]
+O(n-1),
\ee
where $G_9(p)$, $G_{10}(p)$, and $G_{11}(p)$ are defined in (\ref{G9p}), (\ref{G10p}), and (\ref{G11p}), respectively.
Using the analytical continuations (\ref{G9101112}), we obtain contributions from $\varepsilon\varepsilon\varepsilon\varepsilon$
\be \label{F12veveveve}
\cF_{\ve\ve\ve\ve}^{(1/2)} = - \f{1}{8192} ( \lag \ve \rag_\r - \lag \ve \rag_\s )^2
( 19 \lag \ve \rag_\r^2 + 26 \lag \ve \rag_\r \lag \ve \rag_\s + 19\lag \ve \rag_\s^2 ).
\ee

This result for the fermionic theory parallels the bosonic case in the previous subsection.

\section{Free massless compact boson theory} \label{sectionBoson}

The 2D free massless compact boson theory is a CFT with central charge $c=1$. We consider various low-lying primary states, including the ground state $|G\rag$ with conformal weights $(0,0)$, vertex operator state $|\a,\bar\a\rag=|V_{\a,\bar\a}\rag$ with conformal weights $(\f{\a^2}{2},\f{\bar\a^2}{2})$, current state $|J\rag$ with conformal weight $(1,0)$, anti-current state $|\bar J\rag$ with conformal weight $(0,1)$, and state $|J\bar J\rag$ with conformal weights $(1,1)$. Note that the ground state $|G\rag=|0,0\rag$.
For current operators, we have expectation values:
\bea
&& \lag J \rag_{\a,\bar\a}=\f{2\pi\ii\a}{L}, ~~ \lag \bar J \rag_{\a,\bar\a}=-\f{2\pi\ii\bar\a}{L},\\
&& \lag J \rag_J = \lag J \rag_{\bar J} = \lag J \rag_{J\bar J}=
\lag \bar J \rag_J = \lag \bar J \rag_{\bar J} = \lag \bar J \rag_{J\bar J}=0. \nn
\eea
Subsystem fidelities between these states have been obtained in \cite{Lashkari:2014yva,Zhang:2019wqo,Zhang:2019itb}. We classify their short interval expansion calculations into three cases based on the expectation values in these states.

\subsection{Case I} \label{subsection31}

For case I, we have the following exact results
\bea
&& F(\r_{A,G},\r_{A,J})
= F(\r_{A,G},\r_{A,\bar J})
= F(\r_{A,J},\r_{A,J\bar J})
= F(\r_{A,\bar J},\r_{A,J\bar J})
= \f{\G^2(\f{3+\csc\f{\pi x}{2}}{4})}{\G^2(\f{1+\csc\f{\pi x}{2}}{4})} 2\sin(\pi x), \nn\\
&& F(\r_{A,J},\r_{A,\bar J})
= F(\r_{A,G},\r_{A,J\bar J})
= \f{\G^4(\f{3+\csc\f{\pi x}{2}}{4})}{\G^4(\f{1+\csc\f{\pi x}{2}}{4})}
4\sin^2(\pi x),
\eea
where $x = \ell/L$ is the normalized interval length.

Using the universal contributions derived in the previous section, specifically $\cF^{(1/2)}_{\cX\cX}$ (\ref{F12XX}) with $\cX=T,\bar T$ and  $\cF^{(1/2)}_{\cX\cX\cX}$ (\ref{F12XXX}) with $\cX=T,\bar T$, and the factorization formula
\bea \label{FrAsAbosI}
&& F(\r_A,\s_A) = \big[ 1 + \ell^4 \cF^{(1/2)}_{TT}+ \ell^6 \cF^{(1/2)}_{TTT} + O(\ell^8) \big]
\big[ 1 + \ell^4 \cF^{(1/2)}_{\bar T\bar T}+ \ell^6 \cF^{(1/2)}_{\bar T\bar T\bar T} + O(\ell^8) \big] \nn\\
&& \phantom{F(\r_A,\s_A)} = 1 + \ell^4 \big( \cF^{(1/2)}_{TT} + \cF^{(1/2)}_{\bar T\bar T} \big)
+ \ell^6 \big( \cF^{(1/2)}_{TTT} + \cF^{(1/2)}_{\bar T\bar T\bar T} \big)
+ O(\ell^8),
\eea
we reproduce the short interval expansion of the above exact results. This agreement validates our approach and demonstrates that the dominant contributions in these cases come from the stress tensor and its descendants.

To clarify which operators contribute at which order of the short interval expansion for various fidelities, we list the nontrivial leading and subleading terms in Table~\ref{tab} for both the free massless boson theory (discussed in this section) and the free massless fermion theory (discussed in the next section).

\begin{table}
  \centering
  \begin{tabular}{| c | c | c |}
    \hline
      fidelity
    & leading order
    & subleading order \\\hline
    subsections~\ref{subsection31} and \ref{subsection41}
    & $\ell^4 \big( \cF^{(1/2)}_{TT} + \cF^{(1/2)}_{\bar T\bar T} \big)$
    & $\ell^6 \big( \cF^{(1/2)}_{TTT} + \cF^{(1/2)}_{\bar T\bar T\bar T} \big)$ \\\hline
    \multirow{2}{*}{subsections~\ref{subsection32} and \ref{subsection33}}
    & \multirow{2}{*}{$\ell^2 \big( \cF^{(1/2)}_{JJ} + \cF^{(1/2)}_{\bar J\bar J} \big)$}
    & $\ell^4 \big( \cF^{(1/2)}_{JJJJ} + \cF^{(1/2)}_{\bar J\bar J\bar J\bar J}+ \cF^{(1/2)}_{TJJ} + \cF^{(1/2)}_{\bar T\bar J\bar J}$ \\
    &
    & $+ \cF^{(1/2)}_{TJJ} + \cF^{(1/2)}_{\bar T\bar J\bar J}
       + \cF^{(1/2)}_{TT} + \cF^{(1/2)}_{\bar T\bar T}
       + \cF^{(1/2)}_{JJ} \cF^{(1/2)}_{\bar J\bar J} \big)$ \\\hline
    subsections~\ref{subsection42} and \ref{subsection43}
    & $\ell^2 \cF^{(1/2)}_{\ve\ve}$
    & $\ell^4 \big( \cF^{(1/2)}_{\ve\ve\ve\ve} + \cF^{(1/2)}_{T\ve\ve} + \cF^{(1/2)}_{\bar T\ve\ve} + \cF^{(1/2)}_{TT} + \cF^{(1/2)}_{\bar T\bar T} \big)$ \\\hline
  \end{tabular}
  \caption{The nontrivial leading and subleading terms of fidelities in free massless boson and fermion theories. We need the universal contributions $\cF^{(1/2)}_{\cX\cX}$ (\ref{F12XX}) with $\cX=T,\bar T,J,\bar J,\ve$, $\cF^{(1/2)}_{\cX\cX\cX}$ (\ref{F12XXX}) with $\cX=T,\bar T$, $\cF^{(1/2)}_{T\cX\cX}$ (\ref{F12TXX}) with $\cX=J,\ve$, and $\cF^{(1/2)}_{\bar T\cX\cX}$ (\ref{F12TbarXX}) with $\cX=\bar J,\ve$, as well the specific contributions $\cF^{(1/2)}_{JJJJ}$ (\ref{F12JJJJ}), $\cF^{(1/2)}_{\bar J\bar J\bar J\bar J}$ (\ref{F12JbarJbarJbarJbar}) and $\cF^{(1/2)}_{\ve\ve\ve\ve}$ (\ref{F12veveveve}).}
  \label{tab}
\end{table}

\subsection{Case II} \label{subsection32}

For case II, involving the ground state vertex operator states, we have the closed-form expression
\be \label{Fexact}
F(\r_{A,\a,\bar\a},\r_{A,\a',\bar\a'}) = \Big( \cos\f{\pi x}{2} \Big)^{\f{(\a-\a')^2+(\bar\a-\bar\a')^2}{2}}.
\ee

Using the contributions from current operators and stress tensor, $\cF^{(1/2)}_{\cX\cX}$ (\ref{F12XX}) with $\cX=J,\bar J,T,\bar T$,  $\cF^{(1/2)}_{T\cX\cX}$ (\ref{F12TXX}) with $\cX=J$, $\cF^{(1/2)}_{\bar T\cX\cX}$ (\ref{F12TbarXX}) with $\cX=\bar J$, $\cF^{(1/2)}_{JJJJ}$ (\ref{F12JJJJ}), and $\cF^{(1/2)}_{\bar J\bar J\bar J\bar J}$ (\ref{F12JbarJbarJbarJbar}) , and organizing them in a factorized form
\bea \label{FrAsAbosII}
&& F(\r_A,\s_A) = \big[
1 + \ell^2 \cF^{(1/2)}_{JJ}
+ \ell^4 \big(
  \cF^{(1/2)}_{JJJJ}
+ \cF^{(1/2)}_{TJJ}
+ \cF^{(1/2)}_{TT}
\big)
+ O(\ell^6)
\big] \nn\\
&& \phantom{F(\r_A,\s_A) =} \times
\big[
1 + \ell^2 \cF^{(1/2)}_{\bar J\bar J}
+ \ell^4 \big(
\cF^{(1/2)}_{\bar J\bar J\bar J\bar J}
+ \cF^{(1/2)}_{\bar T\bar J\bar J}
+ \cF^{(1/2)}_{\bar T\bar T}
\big)
+ O(\ell^6)
\big] \nn\\
&& \phantom{F(\r_A,\s_A)} = 1 + \ell^2 \big( \cF^{(1/2)}_{JJ} + \cF^{(1/2)}_{\bar J\bar J} \big)
+ \ell^4 \big(
\cF^{(1/2)}_{JJJJ} + \cF^{(1/2)}_{\bar J\bar J\bar J\bar J}
\\&&\phantom{F(\r_A,\s_A)=}
+ \cF^{(1/2)}_{TJJ} + \cF^{(1/2)}_{\bar T\bar J\bar J}
+ \cF^{(1/2)}_{TT} + \cF^{(1/2)}_{\bar T\bar T}
+ \cF^{(1/2)}_{JJ} \cF^{(1/2)}_{\bar J\bar J}
\big)
+ O(\ell^6), \nn
\eea
we reproduce the short interval expansion of the fidelity between vertex operator states. The factorization reflects the holomorphic-antiholomorphic separation in the free boson theory.

\subsection{Case III} \label{subsection33}

For case III, involving mixed states, we have the leading order results
\bea
&& F(\r_{A,J},\r_{A,\a,\bar\a})
= F(\r_{A,\bar J},\r_{A,\bar\a,\a})
= 1 - (\a^2+\bar\a^2) \f{\pi^2 x^2}{16} + o(x^2), \nn\\
&& F(\r_{A,J\bar J},\r_{A,\a,\bar\a}) = 1 - (\a^2+\bar\a^2) \f{\pi^2 x^2}{16} + o(x^2),
\eea
Using the expansion (\ref{FrAsAbosII}), we obtain results with higher-order terms
\bea
&& F(\r_{A,J},\r_{A,\a,\bar\a})
 = F(\r_{A,\bar J},\r_{A,\bar\a,\a})
 = 1 - \f{ (\a^2+\bar\a^2) \pi^2 x^2}{16}
\nn \\ && \phantom{F(\r_{A,J},\r_{A,\a,\bar\a}) = F(\r_{A,\bar J},\r_{A,\bar\a,\a})=}
  + \frac{ [ 3 (\alpha^2+\bar\a^2 )^2+44 \alpha^2-4 \bar\a^2-144 ] \pi^4 x^4}{1536}
  + O(x^6), \\
&& F(\r_{A,J\bar J},\r_{A,\a,\bar\a})
 = 1 - \f{ (\a^2+\bar\a^2) \pi^2 x^2}{16}
  + \frac{ [ 3 (\alpha^2+\bar\a^2)^2+44 (\alpha^2+\bar\a^2 )-288 ]\pi^4 x^4}{1536}
  + O(x^6). \nn
\eea

\begin{figure}[tp]
\centering
\includegraphics[width=0.99\textwidth]{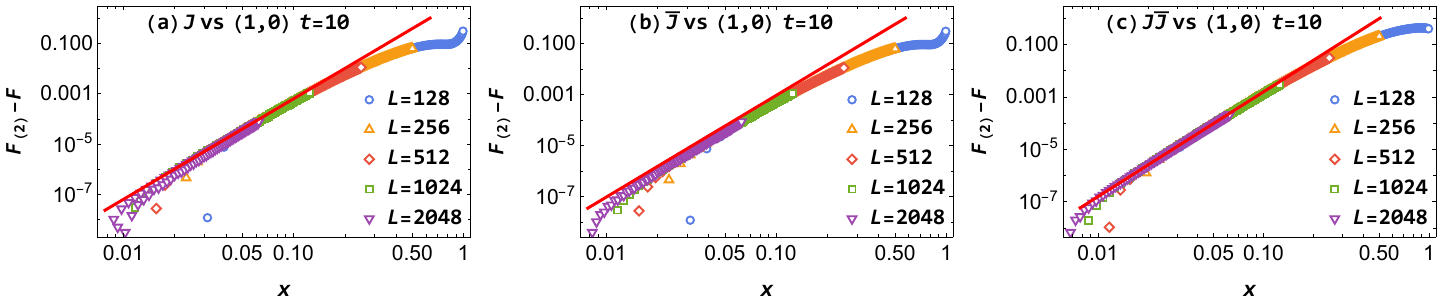}\
\caption{Comparison between CFT predictions (solid red lines) and XX spin chain results (symbols) for the subtracted fidelity $F_{(2)}-F$. In all panels, the spin chain results are obtained from the truncation method with truncation number $t=10$. The good agreement validates the short-interval expansion approach.}
\label{FigureXX}
\end{figure}

\begin{figure}[tp]
\centering
\includegraphics[width=0.99\textwidth]{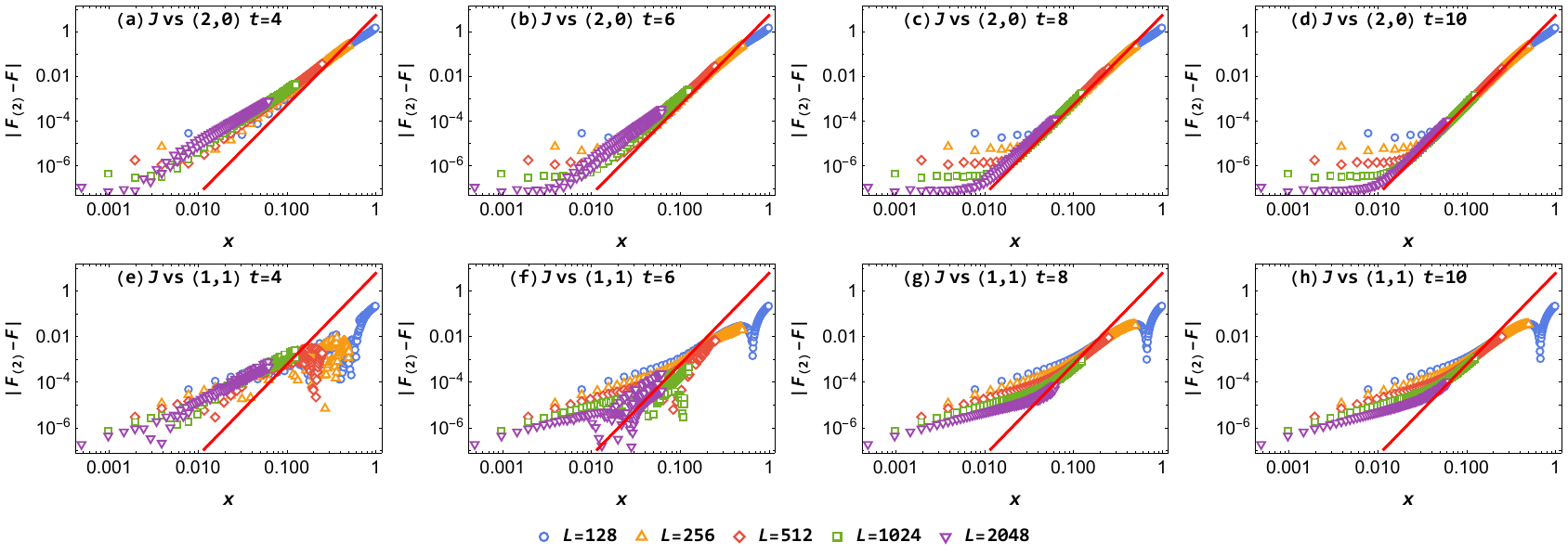}\
\caption{We show additional comparisons of the subtracted fidelity $|F_{(2)}-F|$ between CFT predictions (solid red lines) and XX spin chain results (symbols). The four columns correspond to truncation numbers $t=4,6,8,10$, demonstrating that the numerical results converge to the analytical predictions as $t$ increases.}
\label{FigureXXextra}
\end{figure}

No higher-order analytical results exist in the literature for these cases, so we compare our predictions with numerical results in the spin-1/2 XX chain. In the XX chain, the above CFT results are expected to apply in the limit $1 \ll \ell \ll L$. To match CFT and spin chain results, we require high-efficiency and high-precision numerical evaluation of subsystem fidelity in the spin chain.
We calculate subsystem fidelity in the XX chain using the truncation method with truncation number $t=10$ from \cite{Zhang:2022nuh} and show results in Figure~\ref{FigureXX}. We define the subtracted fidelity
\be
F_{(2)} \equiv 1 - \f{ (\a^2+\bar\a^2) \pi^2 x^2}{16},
\ee
which is removed from both CFT and spin chain results for clearer comparison. The figure uses $(\a,\bar\a)=(1,0)$, showing good agreement between CFT predictions and spin chain results. For other values of $(\a,\bar\a)$, the agreement with theory is not always consistent, likely due to finite-size effects or numerical limitations. Additional results are shown in Figure~\ref{FigureXXextra}. Although the last column does not always show perfect agreement, a clear trend emerges where, with the increase of the truncation number $t$, the numerical results approach the analytical short-interval prediction for $1 \ll \ell \ll L$.

\subsection{Validity range}

\begin{figure}[tp]
\centering
\includegraphics[width=0.66\textwidth]{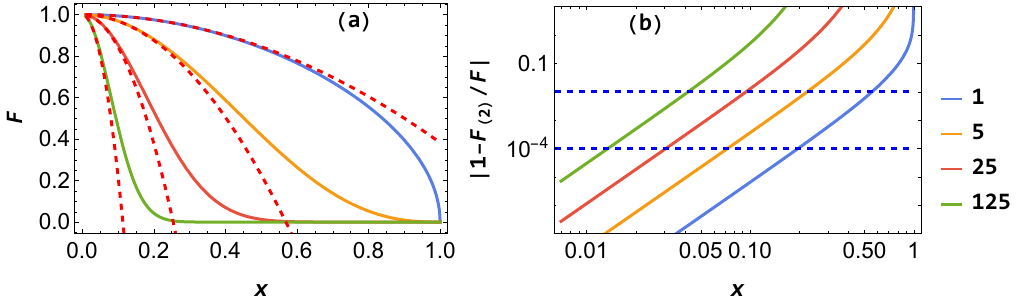}\
\caption{A comparison of the exact fidelity $F(\r_{A,\a,\bar\a},\r_{A,\a',\bar\a'})$ (solid lines, (\ref{Fexact})) and the subtracted fidelity (red dashed lines, (\ref{Fsubtracted})) is shown on the left, with the corresponding relative error on the right. The line colors in both panels represent different values of $(\alpha-\alpha')^2+(\bar\alpha-\bar\alpha')^2$, as shown in the legend.}
\label{FigureFerror}
\end{figure}

The short-interval expansion derived from the OPE of twist operators is expected to be valid only for $\frac{\ell}{L} \ll 1$. A key question is to determine the range of convergence for this expansion and the validity of its finite truncation. This is challenging due to the currently limited number of analytical terms and the precision of our numerical data. Nevertheless, we can analyze this for the known exact analytical results. To this end, we employ the exact fidelity (\ref{Fexact}) and its subtracted fidelity defined as
\begin{equation} \label{Fsubtracted}
F_{(2)} \equiv 1 - \frac{ [ (\alpha-\alpha')^2+(\bar\alpha-\bar\alpha')^2 ] \pi^2 x^2}{16}.
\end{equation}
Figure~\ref{FigureFerror} shows the error of the relative subtracted fidelity, $|1-F_{(2)}/F|$, revealing distinct ranges of validity for different fidelities.

\section{Free massless fermion theory} \label{sectionFermion}

The 2D free massless fermion theory is a CFT with central charge \( c = \frac{1}{2} \). We consider various low-lying primary states, including the ground state \( |G\rangle \) with conformal weights \( (h,\bar{h}) = (0,0) \); the states \( |\sigma\rangle \) and \( |\mu\rangle \), created by the spin operator \( \sigma \) and disorder operator \( \mu \), respectively, both with conformal weights \( (\frac{1}{16},\frac{1}{16}) \); the state \( |\psi\rangle \) with conformal weight \( (\frac{1}{2},0) \); the state \( |\bar{\psi}\rangle \) with conformal weight \( (0,\frac{1}{2}) \); and the state \( |\epsilon\rangle \), created by the energy operator \( \epsilon \), with conformal weights \( (\frac{1}{2},\frac{1}{2}) \).
Expectation values for the energy operator are
\be
\lag \ve \rag_G = \lag \ve \rag_\psi = \lag \ve \rag_{\bar \psi} = \lag \ve \rag_\ve = 0, ~~
\lag \ve \rag_\s = - \lag \ve \rag_\m = \f{\pi}{L},
\ee
We classify related subsystem fidelities between these states from \cite{Zhang:2019wqo,Zhang:2019itb} into three cases.

\subsection{Case I} \label{subsection41}

For case I, we have the exact results
\bea
&& F(\r_{A,G},\r_{A,\psi})
= F(\r_{A,G},\r_{A,\bar \psi})
= F(\r_{A,\psi},\r_{A,\ve})
= F(\r_{A,\bar \psi},\r_{A,\ve})
= \f{\G(\f{3+\csc\f{\pi x}{2}}{4})}
{\G(\f{1+\csc\f{\pi x}{2}}{4})}
\sqrt{2\sin(\pi x)}, \nn\\
&& F(\r_{A,\psi},\r_{A,\bar \psi})
= F(\r_{A,G},\r_{A,\ve})
= \f{\G^2(\f{3+\csc\f{\pi x}{2}}{4})}
{\G^2(\f{1+\csc\f{\pi x}{2}}{4})}
2\sin(\pi x).
\eea
Using the same formula (\ref{FrAsAbosI}) as in the bosonic case, we reproduce the short interval expansion of these results.

\subsection{Case II} \label{subsection42}

For case II, involving the ground state and the spin and disorder states, we have
\bea
&& F(\r_{A,G},\r_{A,\s}) = F(\r_{A,G},\r_{A,\mu}) = \Big( \cos\f{\pi x}{2} \Big)^{\f18}, \nn\\
&& F(\r_{A,\s},\r_{A,\m}) = \Big( \cos\f{\pi x}{2} \Big)^{\f12}.
\eea

From the contributions $\cF^{(1/2)}_{\cX\cX}$ (\ref{F12XX}) with $\cX=\ve,T,\bar T$, $\cF^{(1/2)}_{T\cX\cX}$ (\ref{F12TXX}) with $\cX=\ve$,  $\cF^{(1/2)}_{\bar T\cX\cX}$ (\ref{F12TbarXX}) with $\cX=\ve$, and $\cF^{(1/2)}_{\ve\ve\ve\ve}$ (\ref{F12veveveve}), and using the expansion
\be \label{FrAsAferII}
F(\r_A,\s_A) = 1 + \ell^2 \cF^{(1/2)}_{\ve\ve}
+ \ell^4 \big(
\cF^{(1/2)}_{\ve\ve\ve\ve}
+ \cF^{(1/2)}_{T\ve\ve} + \cF^{(1/2)}_{\bar T\ve\ve}
+ \cF^{(1/2)}_{TT} + \cF^{(1/2)}_{\bar T\bar T}
\big)
+ O(\ell^6),
\ee
we reproduce the short interval expansion of the above results.

\subsection{Case III} \label{subsection43}

For case III, involving mixed states between spin and disorder operators and other primary states, we have the leading order results
\bea
&& F(\r_{A,\s},\r_{A,\psi})
= F(\r_{A,\s},\r_{A,\bar\psi})
= F(\r_{A,\m},\r_{A,\psi})
= F(\r_{A,\m},\r_{A,\bar\psi})
= 1 - \f{\pi^2x^2}{64} + o(x^2), \nn\\
&& F(\r_{A,\s},\r_{A,\ve}) = F(\r_{A,\m},\r_{A,\ve}) = 1 - \f{\pi^2x^2}{64} + o(x^2),
\eea

\begin{figure}
\centering
\includegraphics[width=0.99\textwidth]{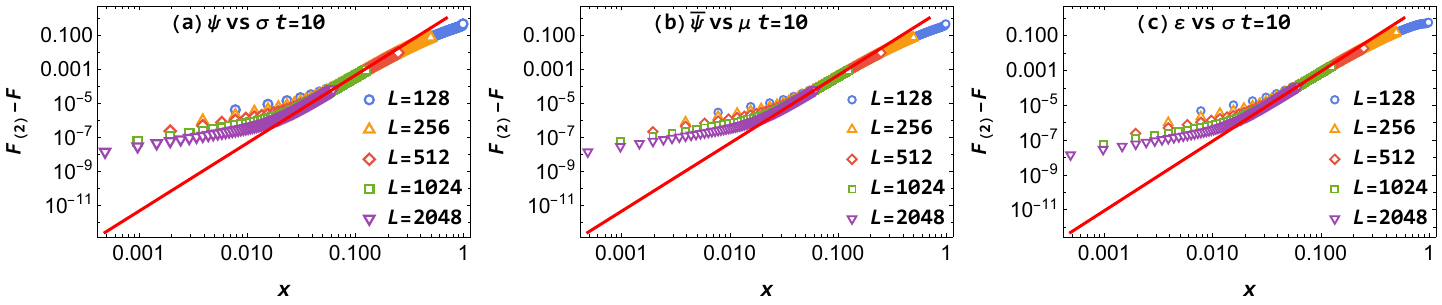}\
\caption{Comparison between CFT predictions (solid red lines) and critical Ising chain results (symbols) for the subtracted fidelity $F_{(2)}-F$. The agreement demonstrates the applicability of our method to fermionic systems.}
\label{FigureIsing}
\end{figure}

Using expansion (\ref{FrAsAferII}), we obtain higher-order corrections
\bea
&& F(\r_{A,\s},\r_{A,\psi})
= F(\r_{A,\s},\r_{A,\bar\psi})
= F(\r_{A,\m},\r_{A,\psi})
= F(\r_{A,\m},\r_{A,\bar\psi})
= 1 - \f{\pi^2x^2}{64} -\frac{1069 \pi^4 x^4}{24576} + O(x^6), \nn\\
&& F(\r_{A,\s},\r_{A,\ve}) = F(\r_{A,\m},\r_{A,\ve}) = 1 - \f{\pi^2x^2}{64} -\frac{2125 \pi^4 x^4}{24576} + O(x^6).
\eea
We compare these predictions with numerical results in the critical Ising chain, obtained using the truncation method from \cite{Zhang:2022nuh} with truncation number $t=10$, as shown in Figure~\ref{FigureIsing}. We define the subtracted fidelity
\be
F_{(2)} \equiv 1 - \f{\pi^2x^2}{64},
\ee
which is removed from both CFT and spin chain results. The agreement between our analytical predictions and numerical simulations provides strong validation of our approach.

\section{2D holographic CFTs} \label{sectionHolographic}

The 2D holographic CFTs have large central charge $c=\f{3R}{2G}$ \cite{Brown:1986nw}, with $G$ being the Newton constant and $R$ being the AdS radius, and sparse low-lying spectrum \cite{Hartman:2014oaa}.
We focus only on contributions from the vacuum conformal family, where holomorphic and anti-holomorphic sectors factorize
\be
F(\r_A,\s_A) = F_{\rm{holo}}(\r_A,\s_A) F_{\rm{anti-holo}}(\r_A,\s_A).
\ee
This section shows only contributions from the holomorphic sector only, as the anti-holomorphic sector follows by analogy. Part of the results in this section has been reported in \cite{Zhang:2025vaq}. Here, we present higher-order results and additional calculation details.

Consider two primary states $|\phi\rag$ and $|\psi\rag$ with large conformal weights $h_\phi = c \e_\phi$ and $h_\psi = c \e_\psi$, respectively. Here $\e_\phi \sim O(c^0)$, $\e_\psi \sim O(c^0)$, $\e_\phi-\e_\psi \sim O(c^0)$. Using our universal formulas (\ref{F12XX}) and (\ref{F12XXX}) with $\cX=T$ and keeping only the contributions from the holomorphic sector of the identity conformal family, we obtain the fidelity expansion
\be \label{FrAphirApsi1}
F(\r_{A,\phi},\r_{A,\psi}) = 1 - \frac{3 \pi^4 c \ell^4 (\epsilon_\phi -\epsilon_\psi )^2}{32 L^4}
+ \frac{\pi^6 c \ell^6 (\epsilon_\phi -\epsilon_\psi )^2 (12 \epsilon_\psi +12 \epsilon_\phi -1)}
{64 L^6}
+ O(\ell^8).
\ee
This result shows that for two heavy states with large different energies in holographic CFTs, the fidelity deviation from unity is of order $O(c)$, reflecting the semiclassical nature of these states in the holographic dual.

Now consider primary states $|\phi\rag$ and $|\psi\rag$ with the same leading order conformal weights $h_\phi = c \e_\phi + \d_\phi$ and $h_\phi = c \e_\phi + \d_\psi$, where $\d_\phi \sim O(c^0)$, $\d_\psi \sim O(c^0)$, and $\d_\phi-\d_\psi \sim O(c^0)$, represent small excitations above a heavy background. In this case, we obtain
\be \label{FrAphirApsi2}
F(\r_{A,\phi},\r_{A,\psi}) = 1 -\frac{3 \pi^4 \ell^4 (\delta_\phi -\delta_\psi )^2}{32 c L^4}
+ \frac{\pi^6 \ell^6 (\delta_\phi -\delta_\psi )^2 [c (24 \epsilon_\phi -1)+12(\delta_\psi+\delta_\phi )]}
{64 c^2 L^6}
+ O(\ell^8).
\ee
At leading order, the two black hole microstates have the same energy and are classically indistinguishable; however, perturbative $1/c$ quantum corrections lift this degeneracy and allow them to be distinguished.

Finally, we consider a primary state $|\phi\rag$ with conformal weight $h_\phi=c\e_\phi$ and a thermal state $\r_\b=\f{\ep^{-\b H}}{Z(\b)}$ with inverse temperature $\b$ and $Z(\b)=\tr(\ep^{-\b H})$. In the case where \cite{Fitzpatrick:2014vua,Fitzpatrick:2015zha}
\be
\b=\f{L}{\sr{24\e_\phi-1}},
\ee
we have $\lag T \rag_\phi=\lag T \rag_\beta$, meaning the states have matching one-point functions of the stress tensor. In this finely tuned situation, we use the our universal formula (\ref{F12XX}) with $\cX=\cA$ and obtain
\be \label{FrAphirAb}
F(\r_{A,\phi},\r_{A,\b}) = 1 - \frac{7 \pi^8 c \ell^8 ( 22 \epsilon_\phi - 1 )^2 \epsilon_\phi^2}{512 (5 c+22) L^8} + O(\ell^{10}).
\ee

In this section, we have only included the contributions from the identity conformal family. If the lightest non-identity operator has conformal weight $\D$, according to (\ref{FrAsA}) its leading contribution to the short interval expansion of the fidelity would be of order $O(\ell^{2\D})$. For the results (\ref{FrAphirApsi1}) and (\ref{FrAphirApsi2}) to be valid, we need $\D>3$, and for the result (\ref{FrAphirAb}) to be valid, we need $\D>4$.

The fidelity (\ref{FrAphirAb}) quantifies how corrections from a quantum theory of gravity encode distinctions between microscopic states into the data accessible to small subsystems \cite{Kudler-Flam:2021rpr, Kudler-Flam:2021alo}. The result directly resolves a puzzle concerning the perturbative distinguishability of black hole microstates. This puzzle, noted in footnote 17 of \cite{Kudler-Flam:2021alo}, stemmed from a tension: studies of fixed-area states, superpositions of energy eigenstates, suggested that distinguishing microstates required non-perturbative precision, whereas evidence from the $1/c$ expansion in holographic CFTs indicated a perturbative effect \cite{Lashkari:2016vgj,Lin:2016dxa,He:2017vyf,Basu:2017kzo,He:2017txy,Guo:2018djz}. The ambiguity arose because the distinguishability of fixed-area states depends on the details of the superposition. Our analysis resolves this by proving that for exact primary states, genuine energy eigenstates dual to black hole microstates, the subsystem trace distance exhibits a universal $1/c$ scaling. This firmly establishes that quantum gravity corrections encode perturbative distinctions between a microstate and the thermal state into small, accessible subsystems.

This demonstrated distinguishability challenges the standard eigenstate thermalization hypothesis (ETH) \cite{Deutsch:2018thx,Dymarsky:2018lhf}, which posits that individual high-energy eigenstates are locally indistinguishable from a thermal ensemble. The violation of ETH in these holographic systems necessitates a generalized description of thermalization. A compelling framework is a generalized Gibbs ensemble (GGE) \cite{Rigol:2006jrd,Cardy:2015xaa} incorporating Korteweg-de Vries (KdV) charges \cite{Bazhanov:1994ft,Cardy:2015xaa,Maloney:2018hdg}, which accounts for additional conserved quantities that distinguish microstates of the same energy. This picture is supported by independent evidence from identity conformal family operators \cite{Dymarsky:2019etq} and large subsystem R\'enyi entropy \cite{Chen:2024lji,Chen:2024ysb}.%

Thus, our proof of perturbative distinguishability for exact eigenstates resolves the foundational puzzle and challenges the standard ETH. While our results necessitate a generalized thermalization picture, the proposal that a GGE with KdV charges universally describes all local observables remains a highly motivated conjecture for future work.%

\section{Conclusion} \label{sectionConclusion}

In this work, we have investigated subsystem fidelity in 2D CFTs using the short-interval expansion derived from the OPE of twist operators. Our approach provides a comprehensive framework for computing fidelity between reduced density matrices of various states in 2D CFTs. We derived explicit expressions for universal contributions from general families of quasiprimary operators, including $\cX\cX$, $\cX\cX\cX$, and $T\cX\cX$, in arbitrary 2D CFTs, capturing both leading and subleading behavior in the short-interval limit. We computed specific contributions from particular quasiprimary operators in free massless boson and fermion theories, demonstrating the method's applicability to concrete examples. Our results show excellent agreement with known analytical expressions and numerical simulations in integrable lattice models, validating the proposed approach. We extended the method to 2D holographic CFTs, analyzing the perturbative distinguishability of black hole microstates through the AdS/CFT correspondence.

The short-interval expansion method developed here offers several advantages: it provides a systematic way to compute fidelity order by order in the interval length, reveals the operator content responsible for state distinguishability. The factorization between holomorphic and antiholomorphic sectors in many cases further simplifies computations and provides physical insight. However, several limitations remain. The expansion is inherently perturbative, and its convergence properties at larger intervals require further investigation. Discrepancies observed in certain cases involving vertex operator states may stem from numerical precision limitations, finite-size effects, or OPE truncation subtleties. Additionally, our holographic treatment relies on the vacuum dominance approximation, which may not capture all significant contributions in finite central charge regimes, while the analytical continuation procedure becomes increasingly complex when higher-order quasiprimary operators are included.

Looking forward, several promising research directions emerge. Extending the short-interval expansion to higher orders by incorporating additional quasiprimary operators would improve accuracy over wider interval ranges. Applying this method to other CFT models, such as supersymmetric theories, non-unitary CFTs, or theories with extended symmetry algebras, could reveal new universal features of subsystem fidelity. Refining the approach for holographic settings may lead to deeper understanding of black hole microstate distinguishability and the information paradox. Furthermore, generalizing the formalism to disjoint intervals, higher-dimensional CFTs, or out-of-equilibrium states would potentially broaden its applicability across quantum information and gravitational physics.

\section*{Acknowledgements}

J.Z. is grateful to the organizers and participants of the Annual Academic Conference of the Gravitation and Relativistic Astrophysics Division of the Chinese Physical Society (Kunming, April 2025), the Symposium on Gravity and Cosmology (Xi'an, June 2025), the Symposium on Strings, Gravity, and Gravitational Waves (Rizhao, August 2025), and the Symposium on Gravity and Cosmology (Hefei, November 2025), where preliminary results of this work were presented and benefitted from valuable discussions. J.Z. also wishes to thank the Institute of Fundamental Physics and Quantum Technology at Ningbo University for its hospitality during a productive visit where part of this research was presented, discussed and completed. In addition, J.Z. thanks Li-Ming Cao, Bin Chen, Anatoly Dymarsky, Wu-zhong Guo, Song He, Yan Liu, Jian-Xin Lu, Jia-Rui Sun, Shao-Jiang Wang, Jie-qiang Wu, and Li-Xin Xu for insightful discussions and constructive comments.

This work is supported by the National Natural Science Foundation of China (Grant No. 12205217) and the Tianjin University Self-Innovation Fund Extreme Basic Research Project (Grant No. 2025XJ21-0007).

\appendix

\section{Analytical continuation} \label{appendixAnalytical}

In the appendix, we collect the analytical continuation for several summation formulas.

We begin by defining
\be \label{G1pD}
G_1(p,\D) = \sum_{a_1<a_2} \f{1}{\big|\sin\f{\pi a_{12}}{2p}\big|^{2\D}},
\ee
where $a_{12}\equiv a_1-a_2$ and $a_1,a_2$ take even integer values in the set $\{0,2,\cdots,2p-2\}$. Our goal is to obtain $G_1(\f12,\D)$ for general $\D$. First, we compute
\be
G_1(p,1)=\frac{p (p^2-1)}{6},
\ee
which gives
\be
G_1\Big(\f12,1\Big)=-\f{1}{16}.
\ee
Proceeding similarly, we obtain the following values
\be
\ba{|c|c|c|c|c|c|c|c|c|c|c|
} \hline
\D & 1 & 2 & 3 & 4 & 5 & 6 & 7 & 8 & 9 & \cdots
\\ \hline
G_1(\f12,\D) & -\frac{1}{16} & -\frac{3}{64} &-\frac{5}{128} & -\frac{35}{1024} & -\frac{63}{2048} &
-\frac{231}{8192} & -\frac{429}{16384} & -\frac{6435}{262144} & -\frac{12155}{524288} & \cdots
\\ \hline
\ea
\ee
From these values, we extract the general formula
\be\label{G112D}
G_1\Big(\f12,\D\Big) = -\frac{\G (\Delta +\frac{1}{2})}{8 \sqrt{\pi } \G(\Delta +1)},
\ee
which can be verified for further values of $\D$.

There is an alternative way to derive the analytical continuation (\ref{G112D}). Using results in \cite{Calabrese:2010he}, we write $G_1(p,\D)$ in (\ref{G1pD}) as
\be
G_1(p,\D) = \f{p(p-1)}{2} g(0) + p \sum_{k=1}^{+\inf} [ p g(p k) - g(k) ],
\ee
with the definition
\be
g(k) = \f{2^{2\D}}{\pi\cos(\pi\D)} \sin[\pi(\D-k)] \int_0^{\pi/2} (\sin\vphi)^{2(\D+k)-1} (\cos\vphi)^{2(\D-k)-1} \dd \vphi.
\ee
Thus we have
\be
G_1\Big(\f12,\D\Big) = -\f18 g(0) + \f12 \sum_{k=1}^{+\inf} \Big[ \f12 g\Big(\f{k}{2}\Big) - g(k) \Big],
\ee
Using
\be
\sum_{k=1}^{+\inf}
\Big\{  \frac{1}{2} \sin \Big[\pi  \Big( \Delta -\frac{k}{2} \Big) \Big] (\tan\varphi)^k
      -\sin [\pi (\Delta -k) ](\tan\varphi)^{2 k}
\Big\}
= -\frac{1}{2} \sin \varphi \cos (\pi  \Delta +\varphi ),
\ee
we get
\bea
&& G_1\Big(\f12,\D\Big) = \f{2^{2\D}}{\pi\cos(\pi\D)} \int_0^{\pi/2}
\Big[
-\f18 \sin(\pi\D)(\sin\vphi)^{2\D-1} (\cos\vphi)^{2\D-1} \\
&& \phantom{G_1\Big(\f12,\D\Big) =}
+\f14 \sin(\pi\D)(\sin\vphi)^{2\D+1} (\cos\vphi)^{2\D-1}
-\f14 \cos(\pi\D)(\sin\vphi)^{2\D} (\cos\vphi)^{2\D}
\Big]
\dd \vphi. \nn
\eea
From
\be
\int_0^{\pi/2} \sin^\a\vphi \cos^\b\vphi \dd \vphi = \f{\G(\f{\a+1}{2})\G(\f{\b+1}{2})}{2\G(\f{\a+\b}{2}+1)},
\ee
which is convergent for $\a>-1$, $\b>-1$, we finally obtain the same analytical continuation (\ref{G112D}). This verifies the method in the previous paragraph. The method in this paragraph relies on special transformations of the expressions, while the method in the previous paragraph is simpler and easier to generalize to more complicated cases, and in the following we will use the simpler method to do the analytical continuations.%

Next, we define
\be \label{G2pD}
G_2(p,\D) = \sum_{a_1<a_2<a_3} \f{1}{\big| \sin\f{\pi a_{12}}{2p} \sin\f{\pi a_{13}}{2p} \sin\f{\pi a_{23}}{2p} \big|^{\D}},
\ee
Using the same method as above, we compute
\be
\begin{tabular}{@{}l@{}}
\begin{tabular}{|c|c|c|c|c|c|c|c|c|}\hline
$\D$ & 2 & 4 & 6 & 8 & 10 & 12 & 14 & 16 \\ \hline
$G_2(\f12,\D)$ & $\frac{3}{128}$ & $\frac{245}{8192}$ & $\frac{12705}{262144}$ & $\frac{2927925}{33554432}$ & $\frac{44757141}{268435456}$ & $\frac{22748036311}{68719476736}$ & $\frac{742182172875}{1099511627776}$ & $\frac{790292983267125}{562949953421312}$ \\ \hline
\end{tabular} \\[+\normalbaselineskip]
\begin{tabular}{|c|c|c|c|c|}\hline
18 & 20 & 22 & 24 & $\cdots$ \\ \hline
$\frac{13349421931027875}{4503599627370496}$ & $\frac{1825735334414506515}{288230376151711744}$ & $\frac{504282766251826384095}{36893488147419103232}$ & $\frac{140415205544153506193175}{4722366482869645213696}$ & $\cdots$ \\ \hline
\end{tabular}
\end{tabular}
\ee
From these results, we extract the general expression
\be \label{G212D}
G_2\Big(\f12,\D\Big) = \frac{2^{\Delta -4} \Gamma (\Delta +\frac{1}{2})^2}
{\pi \Gamma (\frac{\Delta }{2}+1) \Gamma (\frac{3 \Delta }{2}+1)}.
\ee

We now introduce
\be \label{G3pD}
G_3(p,\D) = \sum_{a_1<a_2<a_3} \f{1}{ \big( \sin\f{\pi a_{12}}{2p} \sin\f{\pi a_{13}}{2p} \big)^2
\big| \sin\f{\pi a_{23}}{2p} \big|^{2\D_\cX-2} },
\ee
and obtain the values
\be
\ba{|c|c|c|c|c|c|c|c|c|c|c|
} \hline
\D & 1 & 2 & 3 & 4 & 5 & 6 & 7 & 8 & 9 & \cdots
\\ \hline
G_3(\f12,\D) & \frac{1}{16} & \frac{9}{128} & \frac{35}{512} & \frac{133}{2048}
& \frac{63}{1024} & \frac{957}{16384} & \frac{7293}{131072}
& \frac{27885}{524288} & \frac{26741}{524288} & \cdots
\\ \hline
\ea
\ee
from which we derive
\be \label{G312D}
G_3\Big(\f12,\D\Big) = \frac{(5 \Delta-1) \Gamma (\Delta+\frac{1}{2})}{16 \sqrt{\pi } \Gamma (\Delta+2)}.
\ee

We further define
\be \label{G4pD}
G_4(p,\D) = \sum_{a_1<a_2,b} \f{1}{ \big[ \sin\f{\pi (a_1-b)}{2p} \sin\f{\pi (a_2-b)}{2p} \big]^2
\big| \sin\f{\pi a_{12}}{2p} \big|^{2\D_\cX-2} },
\ee
and compute
\be
\ba{|c|c|c|c|c|c|c|c|c|c|c|
} \hline
\D & 1 & 2 & 3 & 4 & 5 & 6 & 7 & 8 & 9 & \cdots
\\ \hline
G_4(\f12,\D) & -\frac{1}{32} & -\frac{3}{128} & -\frac{5}{256} & -\frac{35}{2048}
& -\frac{63}{4096} & -\frac{231}{16384} & -\frac{429}{32768}
& -\frac{6435}{524288} & -\frac{12155}{1048576} & \cdots
\\ \hline
\ea
\ee
leading to
\be \label{G412D}
G_4\Big(\f12,\D\Big) = -\frac{\Gamma (\Delta+\frac{1}{2})}{16 \sqrt{\pi} \Gamma (\Delta+1)}.
\ee

Next, we consider
\be \label{G5pD}
G_5(p,\D) = \sum_{a_1 \neq a_2,b} \f{1}{ \big[ \sin\f{\pi a_{12}}{2p} \sin\f{\pi (a_1-b)}{2p} \big]^2
\big| \sin\f{\pi (a_2-b)}{2p} \big|^{2\D-2} },
\ee
and find
\be
\ba{|c|c|c|c|c|c|c|c|c|c|c|
} \hline
\D & 1 & 2 & 3 & 4 & 5 & 6 & 7 & 8 & 9 & \cdots
\\ \hline
G_5(\f12,\D) & -\frac{1}{32} & -\frac{3}{64} & -\frac{25}{512} & -\frac{49}{1024}
& -\frac{189}{4096} & -\frac{363}{8192} & -\frac{5577}{131072}
& -\frac{10725}{262144} & -\frac{41327}{1048576} & \cdots
\\ \hline
\ea
\ee
from which we extract
\be \label{G512D}
G_5\Big(\f12,\D\Big) = - \frac{(2 \Delta - 1) \Gamma (\Delta+\frac{1}{2})}{8 \sqrt{\pi } \Gamma (\Delta+2)}.
\ee

We also introduce the definitions
\bea
&& \hspace{-10mm} G_6(p) = \sum_{a_1<a_2<a_3<a_4}
\bigg[
\f{1}{\big( \sin\f{\pi a_{12}}{2p} \sin\f{\pi a_{34}}{2p} \big)^2}
+\f{1}{\big( \sin\f{\pi a_{13}}{2p} \sin\f{\pi a_{24}}{2p} \big)^2}
+\f{1}{\big( \sin\f{\pi a_{14}}{2p} \sin\f{\pi a_{23}}{2p} \big)^2}
\bigg], \label{G6p} \\
&& \hspace{-10mm} G_7(p) = \sum_{a_1<a_2<a_3,b}
\bigg[
\f{1}{\big( \sin\f{\pi a_{12}}{2p} \sin\f{\pi (a_3-b)}{2p} \big)^2}
+\f{1}{\big( \sin\f{\pi a_{13}}{2p} \sin\f{\pi (a_2-b)}{2p} \big)^2}
+\f{1}{\big( \sin\f{\pi (a_1-b)}{2p} \sin\f{\pi a_{23}}{2p} \big)^2}
\bigg], \label{G7p} \\
&& \hspace{-10mm} G_8(p) = \sum_{a_1<a_2,b_1<b_2}
\bigg[
\f{1}{\big( \sin\f{\pi a_{12}}{2p} \sin\f{\pi b_{12}}{2p} \big)^2}
+\f{1}{\big( \sin\f{\pi (a_1-b_1)}{2p} \sin\f{\pi (a_2-b_2)}{2p} \big)^2}
+\f{1}{\big( \sin\f{\pi (a_1-b_2)}{2p} \sin\f{\pi (a_2-b_1)}{2p} \big)^2}
\bigg], \nn \\ \label{G8p}
\eea
with the corresponding analytical continuations
\be \label{G67812}
G_6\Big(\f12\Big) = -\frac{19}{512}, ~~
G_7\Big(\f12\Big) = \frac{3}{128}, ~~
G_8\Big(\f12\Big) = \frac{7}{256}.
\ee

Finally, we define
\bea
&& \hspace{-16mm} G_9(p) = \sum_{a_1<a_2<a_3<a_4}
\bigg(
\f{1}{\sin\f{\pi a_{12}}{2p} \sin\f{\pi a_{34}}{2p}}
-\f{1}{\sin\f{\pi a_{13}}{2p} \sin\f{\pi a_{24}}{2p}}
+\f{1}{\sin\f{\pi a_{14}}{2p} \sin\f{\pi a_{23}}{2p}}
\bigg)^2, \label{G9p} \\
&& \hspace{-16mm} G_{10}(p) = \sum_{a_1<a_2<a_3,b}
\bigg(
\f{1}{ \sin\f{\pi a_{12}}{2p} \sin\f{\pi (a_3-b)}{2p}}
-\f{1}{ \sin\f{\pi a_{13}}{2p} \sin\f{\pi (a_2-b)}{2p}}
+\f{1}{ \sin\f{\pi (a_1-b)}{2p} \sin\f{\pi a_{23}}{2p}}
\bigg)^2, \label{G10p} \\
&& \hspace{-16mm} G_{11}(p) = \sum_{a_1<a_2,b_1<b_2}
\bigg(
\f{1}{\sin\f{\pi a_{12}}{2p} \sin\f{\pi b_{12}}{2p}}
-\f{1}{\sin\f{\pi (a_1-b_1)}{2p} \sin\f{\pi (a_2-b_2)}{2p}}
+\f{1}{\sin\f{\pi (a_1-b_2)}{2p} \sin\f{\pi (a_2-b_1)}{2p}}
\bigg)^2, \label{G11p}
\eea
and obtain the analytical continuations
\be \label{G9101112}
G_9\Big(\f12\Big) = -\frac{19}{512}, ~~
G_{10}\Big(\f12\Big) = \frac{3}{128}, ~~
G_{11}\Big(\f12\Big) = \frac{7}{256}.
\ee

\providecommand{\href}[2]{#2}\begingroup\raggedright\endgroup


\end{document}